\documentclass[runningheads]{llncs}
\usepackage[dvipsnames]{xcolor}
\usepackage{graphicx}
\usepackage{tabularx}
\usepackage{booktabs}
\usepackage{amsmath}
\usepackage{paralist}
\usepackage{todonotes}
\usepackage{hyperref}
\usepackage{algorithm}
\usepackage[noend]{algpseudocode}
\usepackage{microtype}
\usepackage{amssymb}
\usepackage[utf8]{inputenc}
\usepackage[normalem]{ulem}

\RequirePackage{amsmath}

\hypersetup{
	pdftoolbar=true,        %
	pdfmenubar=true,        %
	pdffitwindow=false,     %
	pdfstartview={FitH},    %
	pdftitle={PRIPEL: Privacy-Preserving Event Log Publishing Including Contextual Information},    %
	pdfauthor={},     %
	pdfsubject={},   %
	pdfcreator={},   %
	pdfproducer={}, %
	pdfkeywords={}, %
	pdfnewwindow=true,      %
	colorlinks=true,       %
	linkcolor=Brown,          %
	citecolor=OliveGreen,        %
	filecolor=magenta,      %
	urlcolor=NavyBlue           %
}

\DeclareMathOperator{\ed}{ed}
\DeclareMathOperator{\img}{img}
\DeclareMathOperator{\dom}{dom}

\newcommand{\mypar}[1]{\smallskip\noindent\textbf{#1.}}

\begin{document}
\title{PRIPEL: Privacy-Preserving Event Log Publishing Including Contextual 
Information}

\titlerunning{Privacy-Preserving Event Log Publishing}
\author{Stephan A. Fahrenkrog-Petersen\inst{1} \and Han van der Aa\inst{2} \and Matthias Weidlich\inst{1}
}
\authorrunning{S. Fahrenkrog-Petersen et al.}
\institute{Humboldt-Universität zu Berlin, Berlin, Germany\\
    \email{firstname.lastname@hu-berlin.de}
    \and
    University of Mannheim, Mannheim, Germany\\
    \email{han@informatik.uni-mannheim.de}
}
\maketitle              %
\begin{abstract}
Event logs capture the execution of business processes in terms of executed 
activities and their execution context. Since logs contain potentially 
sensitive information about the individuals involved 
in the process, they should be pre-processed before being published to preserve 
the individuals' privacy. However, existing techniques for such pre-processing 
are limited to a process' control-flow and neglect contextual information, such as attribute values and durations.
This thus precludes any form of process analysis that involves contextual factors.
To bridge this gap,
 we introduce PRIPEL, a framework for privacy-aware event log 
publishing. Compared to existing work, PRIPEL takes a fundamentally different 
angle and ensures privacy on the level of individual cases instead of the 
complete log. This way, contextual information as well as the long tail process 
behaviour are preserved, which enables the application of a rich set of process 
analysis techniques.
We demonstrate the feasibility of our framework in a case study with a 
real-world event log. %

\end{abstract}

\section{Introduction}
\label{sec:introduction}
Process Mining~\cite{van2011process} enables the analysis of business processes 
based on event logs that are recorded by information systems. Events in these 
logs represent the executions of activities as part of a case, including 
contextual information, as illustrated for the handling of patients in an 
emergency room in \autoref{tab:event_log_example}. Such rich event logs do not 
only enable discovery of a model of a process' control-flow, 
see~\cite{augusto2018automated}, but 
provide the starting point for multi-dimensional analysis that incorporates the 
impact of the context on process execution. An example is the prediction of the 
remaining wait time of a 
patient based on temporal information (e.g., arrival in night hours), patient 
characteristics (e.g., age and sex), and activity outcomes (e.g., dispensed 
drugs)~\cite{maggi2014predictive}. 
The inclusion of such contextual information provides a means for 
a fine-granular separation of classes of cases in the analysis. Since 
the separation is largely independent of the frequency of the respective trace 
variants, analysis is not limited to cases that represent common behaviour, but 
includes cases that denote unusual process executions.

Event logs, particularly those that include contextual information, may contain sensitive data related to individuals involved in process 
execution~\cite{MannhardtPO18}. 
Even when explicit pointers to personal information, such as employee names, 
are pseudonymised or omitted from event logs, they remain susceptible to 
re-identification attacks~\cite{garfinkel2015identification}. Such attacks 
still allow personal data of specific individuals to be identified based on 
the contents of an event log~\cite{saskia2019}. Consequently, publishing an 
event log without respective consent violates regulations such as the 
GDPR, 
given that this regulation prohibits processing of personal data for such 
secondary purposes~\cite{voigt2017eu}. This calls for the design of methods 
targeted specifically to protect the privacy of individuals in event logs.
Existing approaches for privacy-preserving process 
mining~\cite{icpm/Fahrenkrog-Petersen19,mannhardt2019privacy} emphasise the 
control-flow dimension, though. They lack the ability to preserve contextual 
information, such as timestamps and attribute values, which 
prevents any fine-granular analysis that incorporates the specifics of 
different classes of cases.
However, aggregations of contextual information in the spirit of 
$k$-anonymity, see~\cite{icpm/Fahrenkrog-Petersen19}, are not suited to 
overcome this limitation. Such aggregations lead to a loss of the 
long tail process behaviour, i.e., infrequent traces of cases that are uncommon 
and, hence, of particular importance for any analysis (e.g., due to exceptional 
runtime characteristics).  
The only existing anonymisation approach that incorporates contextual 
information~\cite{rafiei2018ensuring}  achieves this using homomorphic 
encryption. As such, it fails to provide protection based on any 
well-established privacy guarantee.

\begin{table}[tb]
  \centering
  \scalebox{0.85}{
  \begin{tabularx}{\textwidth}{l l l X}
    \toprule
    Patient ID & Activity & Timestamp &  Payload \\
    \midrule
    2200 & Registration\hspace{1ex} & 03/03/19 23:40:32\hspace{1ex} & \{Age: 
    37, Sex: M,  Arrival:Ambulance\}  \\
    2200 & Triage & 03/05/17 00:47:12 & \{HIV-Positive: True\} \\
    2200 & Surgery & 03/05/17 02:22:17&  \{Operator: House\} \\
    ... & ... & ... & ... \\
    2201 & Registration & 03/05/17 00:01:02 &\{Age: 67, Sex: F,  
    Arrival:Check-In\} \\
    2201 & Antibiotics &03/05/17 00:15:16 & \{Drug: Cephalexin\} \\
    ... & ... & ... & ... \\
    \bottomrule
  \end{tabularx}
}
  \vspace{.5em}
  \caption{Event log example}
  \label{tab:event_log_example}
  \vspace{-3em}
\end{table}

To overcome these gaps, 
this paper introduces \emph{PRIPEL}, a framework for privacy-preserving 
event log publishing that incorporates contextual information. Our idea is to ensure differential privacy of an event 
log on the basis of individual cases rather than on the whole log. To this end, 
the 
PRIPEL framework exploits the maxim of parallel composition of  
differential privacy. Based on a differentially private selection of activity 
sequences, contextual information from the original log is integrated through a 
sequence enrichment step. 
Subsequently, the integrated contextual information is 
anonymised following the principle of local differential privacy. 
Ensuring privacy on the level of individual cases is a fundamentally different 
angle, which enables us to overcome the 
aforementioned limitations of existing 
work. PRIPEL is the first approach to ensure differential privacy not only for 
the control-flow, but also for contextual information in event logs, while 
preserving large parts of the long tail process behaviour. 

Since differential privacy ensures that personal data belonging to specific 
individuals can not longer be identified, the anonymisation achieved by PRIPEL 
is in line with the requirements imposed by the 
GDPR~\cite{wp216,hintze2018viewing}.

We demonstrate the feasibility of our approach through a case study in the 
healthcare domain. Applying PRIPEL to a real-world event log of Sepsis cases 
from a hospital, we show that the anonymisation preserves utility on the level 
of event-, trace-, and log-specific characteristics.

The remainder is structured as follows. In 
\autoref{sec:background}, we provide background in terms of an event log model 
and privacy guarantees. In \autoref{sec:approach}, we introduce the PRIPEL 
framework. 
We present a proof-of-concept in \autoref{sec:proofofconcept}, including an 
implementation and a case study. We discuss our results and reflect on 
limitations in \autoref{sec:discussion}, before we review related work in 
\autoref{sec:relatedwork} and conclude in \autoref{sec:conclusion}.

\section{Background}
\label{sec:background}

This section presents essential definitions and background information. In particular, \autoref{sec:event_log_model} presents the event log model we employ in the paper. Subsequently, \autoref{sec:differential_privacy} defines the foundations of local differential privacy, followed by an introduction to differential privacy mechanisms in \autoref{sec:ensuringprivacy}

\subsection{Event Log Model}
\label{sec:event_log_model}

We adopt an event model that builds upon 
a set of \emph{activities} $\mathcal{A}$. 
An \emph{event} recorded by an information system, denoted by $e$, is assumed 
to be related to the execution of one of these activities, which is written as 
$e.a\in \mathcal{A}$. 
By $\mathcal{E}$, we denote the universe of all events. 
Each event further carries information on its execution context, such as 
the data consumed or produced during the execution of an activity. This payload 
is defined by a set of data attributes $\mathcal{D} = \{ D_1, \ldots, D_p \}$ 
with $\dom(D_i)$ as the domain of attribute $D_i$, $1 \leq i \leq p$. 
We write $e.D$ for the value of attribute $D$ of an event $e$. 
For example, an event representing the activity \emph{`Antibiotics'} may be 
associated with the \emph{`Drug'} attribute that reflects 
the prescribed medication, see \autoref{tab:event_log_example}. Each event $e$
further comes with a timestamp, denoted by $e.ts$, that models the time of 
execution of the respective activity according to some totally ordered time 
domain. %

A single execution of a process, i.e., a case, is represented by a 
\emph{trace}. This is  a sequence $\xi = \langle e_1,\ldots,e_n \rangle$  of events 
$e_i\in \mathcal{E}$, $1 \leq i \leq n$, such that no event 
occurs in more than one trace and the events are ordered by their 
timestamps. We adopt a standard notation for sequences, i.e., $\xi(i)=e_i$ for 
the $i$-th element and $|\xi|=n$ for the length. 
For two distinct traces $\xi = \langle e_1,\ldots,e_n \rangle$ and 
$\xi' = \langle e_1',\ldots,e_m' \rangle$, their concatenation is
$\xi.\xi' = \langle e_1,\ldots,e_n, 
e_1',\ldots,e_m' \rangle$, assuming that the ordering 
is consistent with the events' timestamps. If $\xi$ and $\xi'$ indicate the 
same sequence of activity executions, i.e., 
$\langle e_1.a,\ldots,e_n.a 
\rangle= \langle e_1'.a,\ldots,e_m'.a \rangle$, they are of the same 
\emph{trace variant}. An \emph{event log} is a set of traces, $L = 
\{\xi_1,\ldots,\xi_n \}$, and we write $\mathcal{L}$ for the universe of event 
logs.
\autoref{tab:event_log_example} defines two 
traces, as indicated by the \emph{`patient ID'} attribute. In the remainder, we 
assume the individuals of interest to be represented in at most one case. 
In our example, this means that only 
one treatment per patient is recorded in the log.

\subsection{Foundations of Local Differential Privacy}
\label{sec:differential_privacy}

\emph{Differential privacy} is a definition for privacy that ensures that  personal data of individuals is indistinguishable in a data analysis setting. Essentially, differential privacy aims to allow one to learn nothing about an individual, while learning useful information from a population~\cite{dwork2014algorithmic}.
Achieving differential privacy means that 
 result of a query, performed on an undisclosed dataset, can be published 
 without allowing an individual's personal data to be derived from the 
 published result. On the contrary, methods that achieve \emph{local} 
 differential privacy anonymise a dataset itself in such a manner that it can 
 be published while still guaranteeing the privacy of an individual's 
 data~\cite{kasiviswanathan2011can}. This is achieved by applying noise to  the 
 data,  contrary to applying it to the result of a function performed on 
 the undisclosed data.  The adoption of local differential privacy in industry 
 is well-documented, being employed by, e.g., Apple~\cite{dp2017learning}, 
 SAP~\cite{kessler2019sap}, and 
 Google~\cite{erlingsson2014rappor}.

To apply this notion in the context of event logs, we define $\alpha
: \mathcal{L} \rightarrow \mathcal{L}$
as an \emph{anonymisation function} that takes an event log as input and 
transforms it into an anonymised event log. This transformation is 
non-deterministic and is typically realised through a stochastic function. 
Furthermore, we define $\img(\alpha) \subseteq \mathcal{L}$ as the 
\emph{image} of $\alpha$, i.e., the set of all event logs that may be returned 
by $\alpha$. Finally, we define two event logs $L_1, L_2 \in \mathcal{L}$ to be 
\emph{neighbouring}, if they differ by exactly the data 
of one individual. In our setting, this corresponds to one case and, hence, one 
trace, i.e., $|L_1\setminus L_2| + |L_2\setminus L_1| = 1$. 
Based on~\cite{kasiviswanathan2011can}, we then define local differential 
privacy as follows:

\begin{definition}[Local Differential Privacy]
	\label{def:ldp}
	Given an anonymisation function~$\alpha$ and privacy parameter $\epsilon 
	\in \mathbb{R}$, function $\alpha$ provides $\epsilon$-local 
	differential privacy, if for all neighbouring pairs of event logs $L_1, L_2 
	\in \mathcal{L}$, 
  it holds 
	that:
\begin{equation*}
Pr[\alpha(L_1)\in \img(\alpha)] \leq e^\epsilon \times Pr[\alpha(L_2)\in 
\img(\alpha)] 
\end{equation*}
	where the probability is taken over the randomness introduced by the 
	anonymisation function $\alpha$.
\end{definition}

The intuition behind the guarantee is that it limits the information that can 
be disclosed by one individual, i.e., one case. 
The strength of the guarantee depends on 
$\epsilon$, with lower values leading to  stronger data protection.

\subsection{Ensuring Local Differential Privacy}
\label{sec:ensuringprivacy}

Mechanisms that ensure local differential privacy strive to provide privacy 
guarantees while keeping as much useful information as possible, i.e., they aim 
to maintain maximum utility of the dataset. The mechanisms typically do not 
delete or generalize (parts of the) data, as is done to obtain other privacy 
guarantees~\cite{lefevre2005incognito}. Rather, they define an anonymisation 
function that inserts noise into 
data, in order to obscure information about individuals, while retaining as many 
characteristics about the general population as possible. Several such 
mechanisms have been developed to anonymise various data types, including ones that ensure differential 
privacy for numerical, categorical, and boolean data: 

\mypar{Numerical data -- Laplace mechanism} 
The Laplace mechanism~\cite{dwork2006calibrating} is an additive noise 
mechanism for numerical values. It draws noise from a Laplacian 
distribution, that is calibrated based on the privacy parameter $\epsilon$ and 
the sensitivity of the data distribution. %
The latter is defined as the maximum difference one individual can cause.

\mypar{Boolean data - Randomized response}  To ensure differential privacy of 
boolean data, one can use \emph{randomized 
response}~\cite{warner1965randomized}. The algorithm is based on the following 
idea: A fair coin toss determines if the true value of an individual is 
revealed or if a randomized value is chosen instead. Here, the randomization 
depends on the strength $\epsilon$ of the differential privacy guarantee. In 
this paper, we will use a so-called %
 \emph{binary mechanism}~\cite{holohan2017optimal}.

\mypar{Categorical data - Exponential mechanism}
To handle categorical data, it is possible to use the \emph{exponential 
mechanism}~\cite{mcsherry2007mechanism}. It enables the definition of a utility 
difference between the different potential values of the domain of the 
categorical value. The probability of a value being exchanged by another value 
depends on the introduced probability loss.

\mypar{Parallel composition of differential privacy}
Given such mechanisms that are able to provide differential privacy for various 
data types, a  crucial property of (local) differential privacy is that it is 
compositional.
Intuitively, this means that when the results of multiple 
$\epsilon$-differential-private mechanisms, performed on disjoint datasets,  
are merged, the merged result also provides $\epsilon$-differential 
privacy~\cite{mcsherry2009privacy}. 
Adapted to our notion of attributes and timestamps of events, this is 
formalized as follows: Let $M_i(e.d_i)$, $1 \leq i \leq p$, and $M_0(e.ts)$ be 
the values obtained by some mechanisms $M_0, M_1, \ldots M_p$ for the attribute 
values and the timestamp of 
an event $e$. Then, if all mechanisms provide $\epsilon$-differential privacy 
and under the assumption of all attributes (and the timestamp) being 
independent, 
the result of their joint application to $e$ also provides 
$\epsilon$-differential privacy.

This property forms a crucial foundation for our proposed framework to 
privacy-aware event log publishing, as introduced next.

\section{The PRIPEL Framework}
\label{sec:approach}
\begin{figure}[!htb]
	\centering
	\includegraphics[width=.8\columnwidth]{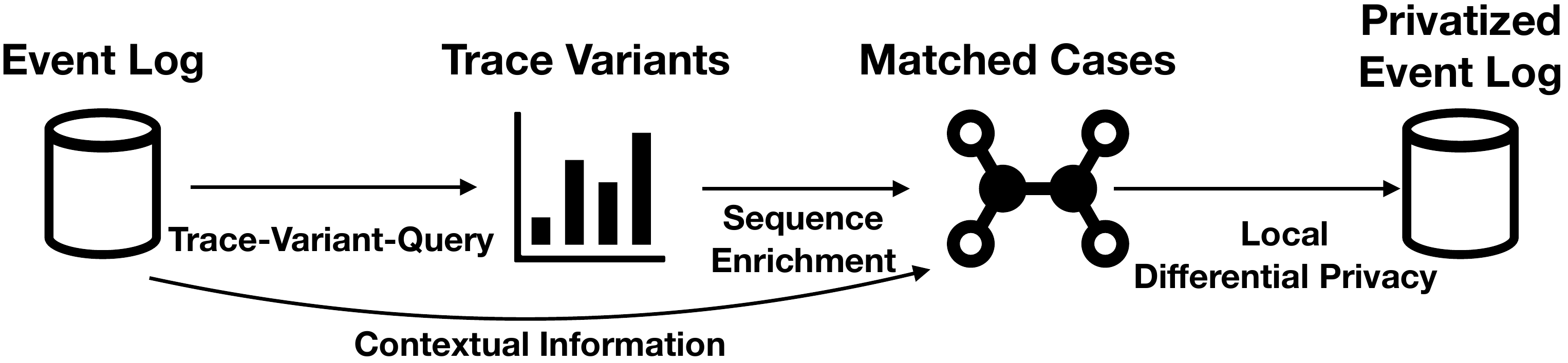}
	\caption{Overview of PRIPEL Framework}
	\label{fig:framework}
  \vspace{-1.5em}
\end{figure}

The \emph{Pri}vacy-\emph{P}reserving \emph{e}vent \emph{l}og publishing 
(\emph{PRIPEL}) framework takes an event log as input and transforms it into an 
anonymised one that includes contextual information and guarantees 
$\epsilon$-differential privacy. 
As depicted in \autoref{fig:framework}, the PRIPEL framework consists of 
three main steps. 
Given an event log~$L$, PRIPEL first applies a \emph{trace-variant query} $Q$ 
on $L$. The query returns a bag of activity sequences that ensures 
differential privacy from a control-flow perspective. 
Second, the framework constructs new traces by enriching the activity sequences 
obtained by $Q$ with contextual information, i.e., timestamps and attribute 
values, from the original log $L$. This is achieved in a \emph{sequence 
enrichment} step, which results in a \emph{matched} event log $L_m$. 
Finally, PRIPEL anonymises the timestamps and attribute values 
of $L_m$ individually by exploiting the maxim of parallel composition of 
differential privacy. 
The resulting event log $L'$ then guarantees $\epsilon$-differential privacy, 
while largely retaining the information of the original log~$L$.

Sections~\ref{sec:trace_variant_query} 
through~\ref{sec:local_differential_privacy} outline instantiations of each of 
these three steps. However, we note that the framework's steps can also be 
instantiated in a different manner, for instance by using alternative 
trace-variant queries or matching techniques. It is therefore possible to 
tailor PRIPEL to specific use cases, such as a setting in which traces become 
available in batches. 

\subsection{Trace Variant Query}
\label{sec:trace_variant_query}

The first step of our framework targets the anonymisation of an event log from 
a control-flow perspective. In particular, the framework applies a 
trace variant query, which returns a bag of activity sequences that 
captures trace variants and their frequencies in a differentially private 
manner. Such a step is essential, given that even the publication of activity 
sequences from an event log, i.e., with all attribute values and timestamps 
removed, can be sufficient to link the identity of individuals to infrequent 
activity 
sequences~\cite{icpm/Fahrenkrog-Petersen19,mannhardt2019privacy}. For example, 
uncommon treatment paths may suffice to resolve the identity of a specific 
patient.

In PRIPEL, we adopt a state-of-the-art realisation of a privacy-preserving 
trace variant query~\cite{mannhardt2019privacy}. It employs a Laplace mechanism 
(see \autoref{sec:ensuringprivacy}) to add noise to the result of a trace 
variant query. As shown for an exemplary query result in 
\autoref{tab:example_trace_variant_query}, this mechanism may alter the 
frequency of trace variants, remove variants entirely, and introduce new ones. 
Note that the size of a trace variant query typically differs from 
the number of traces in the original log.

The employed trace variant query is configured with two  
parameters, $n$ and~$k$, which influence the prefix-tree that the mechanism 
uses to generate a query result. 
Here, $n$ sets the maximum depth of the prefix-tree, which determines the 
maximum length of an activity sequence returned by the query. 
Parameter~$k$ is used to bound the mechanism's state space in terms of the 
number of potential activity sequences that are explored. 
A higher~$k$ means that only more commonly occurring prefixes are considered, 
which reduces the runtime, but may negatively affect the resulting log's 
utility.
The runtime complexity of the query depends on the maximal number of explored 
prefixes: $\mathcal{O}(|\mathcal{A}|^n)$. Yet, in practice, the exponential 
runtime is mitigated by the pruning parameter~$k$.

\begin{table}[h]
  \vspace{-1.2em}
  \begin{tabularx}{\textwidth}{l @{\hspace{1em}} X X}
    \toprule
    Trace Variant & Count & Privatized Count \\
    \midrule
    $\langle \mathit{Registration},\mathit{Triage},\mathit{Surgery}\rangle$ & 5 
    & 
    6\\
    $\langle \mathit{Registration},\mathit{Triage},\mathit{Antibiotics}\rangle$ 
    & 
    7 & 5\\
    $\langle \mathit{Registration},\mathit{Triage},\mathit{Surgery}, 
    \mathit{Antibiotics}\rangle$ & 2 &  3\\
    $\langle 
    \mathit{Registration},\mathit{Triage},\mathit{Antibiotics},\mathit{Surgery},\mathit{Antibiotics}\rangle$
    & 
    0 &  1\\
    \bottomrule
  \end{tabularx}
  \caption{Illustration of a privacy-aware trace variant query}
  \label{tab:example_trace_variant_query}
  \vspace{-2.5em}
\end{table}

Below, we adopt a flattened representation of the result of the 
trace variant query. By $Q(L)\subseteq (\mathcal{A}^*)^*$, we denote a 
sequence of activity sequences derived by duplicating each activity sequence 
returned by the trace variant query according to its frequency, in 
some arbitrary order. For example, if the query returns the bag 
$[\langle 
\mathit{Registration},\mathit{Triage}\rangle^2,
\langle \mathit{Registration},\mathit{Triage},\mathit{Antibiotics}\rangle]$, 
$Q(L)$ is defined as 
$\{\langle \mathit{Registration},\mathit{Triage}\rangle,
\langle \mathit{Registration},\mathit{Triage},\mathit{Antibiotics}\rangle, 
\langle \mathit{Registration},\mathit{Triage}\rangle
\}$.

So far, no other designs for trace variant queries have been introduced in the 
literature. However, we assume that 
alternative query formulations suited for specific use cases will be developed 
in the future.

\subsection{Sequence Enrichment}
\label{sec:trace_matching}

The second step of the framework enriches the activity sequences obtained 
by the trace variant query with 
contextual information, i.e., with timestamps and attribute values. This is 
achieved by establishing a trace matching between each activity sequence 
from~$Q(L)$ and a trace of the original log~$L$. The latter trace determines 
how the activity sequence is enriched with contextual information to construct 
a trace of the matched log~$L_m$. Here, $L_m$ should resemble the 
original log: 
Distributions of attribute values and timestamps, along with  
their correlation with trace variants in the original $L$ shall be mirrored 
in the matched log~$L_m$. 

To link the activity 
sequences in $Q(L)$ and traces in log $L$, we define a matching 
function $f_m: Q(L) \nrightarrow L$. It is potentially partial and injective, 
i.e., it matches 
each activity sequence (again, note that activity sequences obtained from the 
trace variant query are duplicated according to their frequency) to a separate 
trace in $L$, 
such that $f_m(\sigma_1) = f_m(\sigma_2)$ 
implies that $\sigma_1=\sigma_2$ for all $\sigma_1, \sigma_2 $ that are part of 
$Q(L)$. 
However, constructing such a mapping function requires to address two 
challenges:
\begin{compactenum}[(i)]
	\item Since the trace variant query introduces noise, some sequences 
	from $Q(L)$ cannot be paired with traces in $L$ that are of the exact 
	same sequence of activity executions. Given a sequence 
	$\sigma = \langle 
	\mathit{Registration},\mathit{Triage},\mathit{Release} \rangle$ of $Q(L)$ 
	and a trace 
	$\xi$ with its activity executions being $\langle 
	\mathit{Registration},\mathit{Release}\rangle$, for example, the trace does 
	not provide 
	attribute values to be assigned to a \emph{`Triage'} event. To preserve their order, the 
	insertion of an additional event may require the timestamps of other events to be changed as well.
	
	\item Since $Q(L)$ may contain more sequences than the original log $L$ has 
	traces, some sequences in $Q(L)$ might not be matched to any trace in $L$, 
	i.e., $f_m$ is partial. 
	Since all sequences in $Q(L)$ must be retained in the construction of traces 
	for the matched log to ensure differential privacy, also such 
	\emph{unmatched} sequences must be enriched with contextual information. 
\end{compactenum}

\noindent Given these challenges, PRIPEL incorporates three 
functions: (1) a matching function $f_m$; (2) a mechanism $f_e$ to enrich 
a matched sequence $\sigma$ with contextual information from trace 
$f_m(\sigma)$ to 
construct a trace 
for the matched log $L_m$; and 
(3) a mechanism $f_u$ to enrich an unmatched sequence to construct a
trace for~$L_m$. In this paper, we propose to instantiate these functions as 
follows:

\mypar{Matching function}
The matching function $f_m$ shall establish a mapping from $Q(L)$ to $L$ such 
that the activity sequences and traces are as \emph{similar} as possible. This 
similarity can be quantified using a distance function. Here, we propose to use 
the Levensthein distance~\cite{levenshtein1966binary} to quantify the edit 
distance of some sequence $\sigma$ that is part of $Q(L)$ and the sequence of 
activity executions derived from a trace $\xi \in L$, denoted as $\ed(\sigma, 
\xi)$.
Using assignment optimization techniques, the matching function is 
instantiated, such that the total edit distance is minimized, i.e., with 
$Q(L) = \langle \sigma_1,\ldots, \sigma_n\rangle$, we minimize 
$\sum_{1\leq i\leq n} 
\ed(\sigma_i, f_m(\sigma_i))$.

\begin{algorithm}[t]%
	\caption{Matched Sequence Enrichment}
	\label{alg:resolve_trace_matching}
	\begin{flushleft}
		\textbf{INPUT:} An event log $L$; an activity sequence $\sigma$; the 
		matched trace $\xi = f_m(\sigma)$.\\
		\textbf{OUTPUT:} A trace $\xi_{\sigma}$ derived by enriching $\sigma$ based 
		on $\xi$.
		\vspace{-.6em}
	\end{flushleft}
	\begin{algorithmic}[1]
		\For{$1\leq i\leq |\sigma|$}
		\State $e \gets $ create new event  %
		\State $e.a \gets \sigma(i).a$ \Comment{Assign activity to new event}
		\State $k_{\sigma} \gets |\{ 1\leq j \leq |\xi_{\sigma}| \mid 
		\xi_{\sigma}(j).a = 
		e.a\}|$ 
		\Comment{Count $a$-events in new trace $\xi_{\sigma}$} 
		\label{line:nthoccurrence}
		\State $k_{\xi} \gets |\{ 1\leq j \leq |\xi| \mid \xi(j).a = 
		e.a\}|$ 
		\Comment{Count $a$-events in original trace $\xi$} 
		\If{$k_{\sigma} < k_{\xi}$} 
		\Comment{Get corresponding occurrence of $a$} 
		\State $e' \gets 
		\xi(j)$ with $\xi(j).a=e.a$ and $|\{ 1\leq l < j \mid \xi(l).a = 
		e.a\}| = k_{\sigma}$
		\label{line:getnthoccurrence}
		\ForAll{$D \in \mathcal{D}$} 
		\State $e.D \gets e'.D$
		\Comment{Assign attribute values of $e'$ to  $e$} 
		\EndFor
		\label{line:assignattributes}
		\If{$e'.ts > \xi_{\sigma}(|\xi_{\sigma}|).ts$} 		
		\label{line:check_timestamp}
		\State $e.ts \gets e'.ts$  \label{line:assign_timestamp}
		\Else \ 
		\State $e.ts \gets $ derive timestamp based on \autoref{eq:timestamp} 
		\label{line:random_timestamp}
		\EndIf
		\Else \Comment{No corresponding event in $\xi$}\label{line:no_event}
		\ForAll{$D \in \mathcal{D}$} 
		$e.D \gets$ draw random attribute value
		\label{line:random_event}
		\EndFor
		\State $e.ts \gets $ draw random timestamp for activity $e.a$ 
		\label{line:random_timestamp_2}
		\EndIf
		\State $\xi_{\sigma} \gets \xi_{\sigma} . \langle e\rangle$ 
		\EndFor
		\State \Return $\xi_{\sigma}$ \label{line:return} \Comment{Return new trace}
	\end{algorithmic}
\end{algorithm}

\mypar{Matched sequence enrichment}
Given a matched sequence $\sigma$ of $Q(L)$, the sequence $\sigma$ is 
enriched based on the context information of trace $\xi=f_m(\sigma)$ to create 
a new trace $\xi_{\sigma}$. 
The proposed procedure for this is described by \autoref{alg:resolve_trace_matching}. 
To create the events for the new trace $\xi_{\sigma}$ derived from $\sigma$, we 
iterate over all activities in $\sigma$, create a new event, and check if there 
is a corresponding event~$e'$ of~$\xi$. Using $k_\sigma$ as the number of times 
we have observed activity~$a$ in the 
sequence~$\sigma$ (line~\ref{line:nthoccurrence}), $e'$ shall be the 
$k_\sigma$-th occurrence of an event in $\xi$ with $e.a = 
a$ (line~\ref{line:getnthoccurrence}). 
If such an event $e'$ exists, we assign all its attribute values to the new 
event $e$ (line~\ref{line:assignattributes}). 
Subsequently, we check if the timestamp of $e'$ occurs after the 
timestamp of the last event of $\xi_{\sigma}$ 
(line~\ref{line:check_timestamp}). 
If this is the case, we assign the timestamp $e'.ts$ of the original event to 
event $e$. Otherwise, we generate a new timestamp based on the following 
equation, assuming that the current event is the $n$-th event to be added to 
$\xi_\sigma=\langle e_1,\ldots,e_{n-1}  \rangle$: 
\begin{equation}
\label{eq:timestamp}
e_{n}.ts = e_{n-1}.ts + \Delta_{e_{n-1}.a,e_{n}.a}
\end{equation}
Here, $\Delta_{e_{n-1}.a,e_{n}.a}$ denotes a timestamp difference randomly 
drawn from the distribution of these differences in the original log. That is, 
the distribution is obtained by considering all pairs of subsequent events in 
the original traces that indicate the execution of the respective activities.  
If no such pairs of events appeared in the original log, we resort to the 
distribution of all timestamp differences of all pairs of subsequent activities 
of the original log.

If no corresponding event $e'$ can be found for the newly created event $e$, we 
assign randomly drawn attribute values and a timestamp to this event 
(lines~\ref{line:random_event}--\ref{line:random_timestamp_2}). We draw the 
attributes values from the overall distribution of each attribute $D$ 
in the original log $L$, while timestamps are calculated according to 
\autoref{eq:timestamp}.

\mypar{Unmatched sequence enrichment} For sequences in $Q(L)$ without a 
matching, we assign the attribute values randomly. To handle the timestamps, we 
randomly draw a timestamp $t_{start}$ for the event created for the first 
activity in $\sigma$, from the overall distribution of all timestamps of 
the first events of all traces $\xi$ in the original log $L$. We generate the 
remaining timestamps based on \autoref{eq:timestamp}.

The runtime complexity of the whole sequence enrichment step is dominated by 
the assignment optimization problem, which requires $\mathcal{O}(|Q(L)|^3)$ 
time.

\subsection{Applying Local Differential Privacy}
\label{sec:local_differential_privacy}

Next, starting with the matched log derived in the previous step, we turn to 
the anonymisation of contextual information using local differential privacy. 
While the treatment of attribute values follows rather directly from existing 
approaches, we propose a tailored approach to handle timestamps. The runtime 
complexity of this step is linear in the size of the matched log $L_m$, i.e., 
we arrive at $\mathcal{O}(|L_m|)$.

\mypar{Anonymising attribute values} We differentiate between attributes of 
three data types: numerical, categorical, and boolean. For each type, we employ 
the mechanism discussed in \autoref{sec:ensuringprivacy}. 
Under the aforementioned assumptions for parallel composition of differential 
privacy, the resulting values are $\epsilon$-differentially private.
Note that for each attribute, a different privacy parameter 
$\epsilon$ may be chosen. This way, the level of protection may be adapted to 
the sensitivity of the respective attribute values.

\begin{figure}[t]
  \centering
  \includegraphics[width=.9\columnwidth]{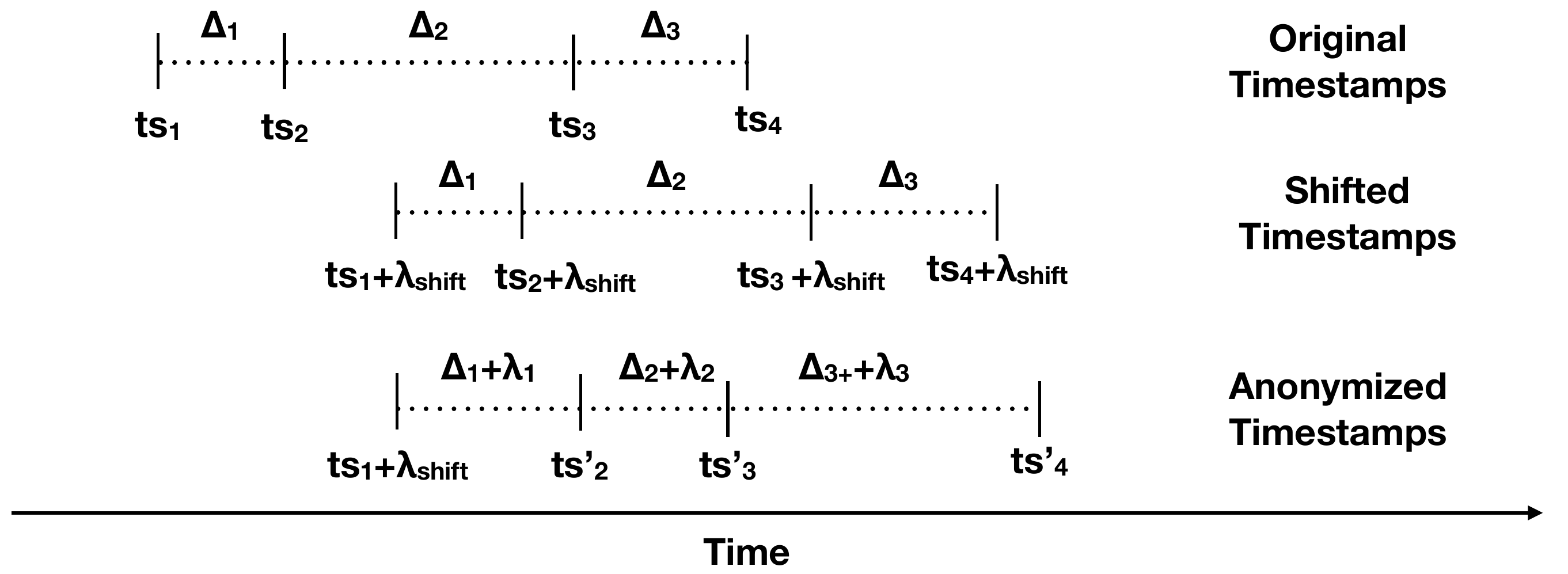}
  \vspace{-.5em}
  \caption{Illustration of timestamp anonymisation}
  \label{fig:timestamp_anonymisation}
  \vspace{-1em}
\end{figure}

\mypar{Anonymising timestamps} To anonymise timestamps, we introduce random 
timestamp shifts, which is inspired by the treatment of network 
logs~\cite{zhang2006outsourcing}.
That is, we initially alter all timestamps based on some randomly drawn noise 
value, $\lambda_{shift}$, which is drawn, for instance, from a Laplacian  
distribution. The result is illustrated in the  middle sequence of 
\autoref{fig:timestamp_anonymisation}. 
After this initial shift, we subsequently introduce noise to the time intervals 
between events, depicted as $\Delta_1$, $\Delta_2$, and $\Delta_3$ in the 
figure. To this end, we add random noise to the length of each interval, 
denoted by $\lambda_1$, $\lambda_2$, and $\lambda_3$.
To retain the order of events, we bound the random timestamp shift to the size 
of the interval between two events. Since the event order was already 
anonymised in the first step of the framework 
(\autoref{sec:trace_variant_query}), introducing additional noise by 
re-ordering events here would just reduce the event log's utility.

After this final step, all aspects of the original log, i.e., control-flow 
and contextual information, have been 
anonymised. Based on the maxim of parallel composition, the resulting log 
provides $\epsilon$-differential privacy. 

\section{Proof-of-Concept}
\label{sec:proofofconcept}
This section presents a proof-of-concept of the PRIPEL framework. 
We first report on a prototypical implementation 
(\autoref{sec:implementation}), which we apply in a case study using a 
real-world event log 
(Sections~\ref{sec:casestudysetup}--\ref{sec:casestudyresults}).
In this manner, we aim to show the feasibility of the framework in 
a realistic setting and investigate its ability to preserve the utility of an 
event log while providing privacy guarantees.

\subsection{Prototypical Implementation}
\label{sec:implementation}
We implemented PRIPEL in Python and published our implementation under the MIT 
licence on Github.\footnote{\url{https://github.com/samadeusfp/PRIPEL}} 
The implementation uses the PM4Py library~\cite{berti2019process} to parse and 
process event logs. To instantiate the framework, we implemented a Python 
version of the trace-variant query by Mannhardt et 
al.~\cite{mannhardt2019privacy}. The anonymisation of contextual information is 
based on IBM's \emph{diffprivlib} library~\cite{holohan2019diffprivlib}.

\subsection{Case Study Setup}
	\label{sec:casestudysetup}

We show the feasibility of PRIPEL by applying our implementation to the Sepsis 
event log~\cite{mannhardt2016sepsis}. We selected this event log given its 
widespread adoption as a basis for case studies, as well as due to  the 
relevance of its characteristics in the context of our work. As shown in our 
earlier work~\cite{icpm/Fahrenkrog-Petersen19}, anonymisation techniques that 
perform aggregations over the whole Sepsis log have a considerable impact on 
the anonymised log's utility. The reason being the long tail process behaviour 
in terms of a relatively low number of re-occurring trace variants: 1,050 
traces spread over 846 trace variants. As such, the log's challenging   
characteristics make it particularly suitable for a proof-of-concept with our 
framework.

To parametrise our implementation, we test different values of the privacy 
parameter $\epsilon$, ranging from 0.1 to 2.0. 
Given that this parameter defines the strictness of the desired privacy guarantees (lower being stricter), varying $\epsilon$  shall show its impact on utility of the resulting anonymised log.

We select the maximal prefix length $n=30$, to cover the length of over 95\% of 
the 
traces in the Sepsis event log. To cover all potential 
prefixes of the original log, we would need to set $n=185$. However, this would 
add a lot of noise and increase the runtime significantly. Therefore, we opt 
for only looking into shorter traces.
For each event log, we opted for the lowest value for $k$ that still computes 
the query within a reasonable time, as will be detailed in the remainder.

\subsection{Case Study Results}
\label{sec:casestudyresults}

In this section, we first focus on the runtime of the PRIPEL framework. 
Subsequently, we explore its ability to preserve event log utility while 
guaranteeing $\epsilon$-differential privacy.

	\mypar{Runtime} We measured the runtime of our PRIPEL implementation for 
	various parameter configurations, obtained on a MacBook Pro (2018) with an 
	i5 Intel Core CPU and 8GB memory. As shown in \autoref{tab:runtime}, we 
	were typically able to obtain an anonymised event log in a manner of 
	minutes, which we deem feasible in most application scenarios.
	However,  the runtime varies considerably across the chosen configurations 
	and the framework's three main steps. 

	All besides one of the anonymised logs have far more traces than the 
	original log, due to the added noise as part of the trace variant query. 
	However, this is not true for the log with a  
	$\epsilon=1.5$ differential privacy guarantee, which contains only one 
	third of the number of traces of the original log. This is due to the low 
	noise level and the fact that $k=2$ cuts out all variants that appear only 
	once. 
	This applies to nearly all the variants in the original log. Since only a 
	few noisy traces are added, the resulting log is significantly smaller than 
	the original log.
	
\begin{table}
\vspace{-1em}
\centering
\setlength{\tabcolsep}{8pt}
\begin{tabular}{c c r rrrr}
	\toprule
	$\epsilon$ &$k$ & $|Q(L)|$ & Query & Enrichment&Anonymisation & 
	Total \\
	\midrule
	0.1  & 20 & 5,175 &1s & 35s  & 3m24s & 4m07s  \\
	0.5 &4 & 6,683 &1s & 3m52s  & 4m08s & 8m12s  \\
	1.0 &2 & 7,002 & 2s & 8m37s  & 4m27s & 13m18s  \\
	1.5  &2 & 340 & 1s & 8s  & 13s & 23s  \\
	2.0 &1 & 13,152 & 9s & 33m05s  & 8m30s & 42m06s \\
	\bottomrule
\end{tabular}
\vspace{.7em}
\caption{Runtime of \emph{PRIPEL} for the Sepsis log}
\label{tab:runtime}
\vspace{-2em}
	\end{table}

The trace variant query (Step 1 in PRIPEL), is executed in a manner of seconds, 
ranging from one to nine seconds, depending on the configuration. However, this 
runtime 
could be greatly exceeded for configurations with a higher $n$. 
While a trace variant query with $\epsilon=1.5$ and $k=2$ is answered in one 
second, a configuration of $\epsilon=1.5$ and $k=1$ does not lead to any result 
within an hour.

Sequence enrichment (Step 2) is the step with the largest runtime 
variance, from 35 seconds to 33 minutes. In most configurations, this step also 
represents the largest contribution to the total runtime. 
This is due to the polynomial runtime complexity of the enrichment step, see
\autoref{sec:trace_matching}. To reduce this runtime, a 
greedy strategy may instead be used to match activity sequences and traces.

Anonymisation based on local differential privacy (Step 3) has a reasonable 
runtime that increases linearly with the number of traces in the resulting log. 

Based on these observations and the non-repetitive character of the 
anonymisation task, we argue that it is feasible to apply our PRIPEL 
framework in real-world settings. However, if runtime plays a crucial 
factor in an application scenario, it should be clear that a suitable 
parameter configuration must be carefully selected.

\mypar{Event log utility} To illustrate the efficacy of PRIPEL, we analyse the 
utility of anonymised event logs. In particular, we explore measures for three 
scopes: (1) the event level, in terms of \emph{attribute value quality}, (2) 
the trace level, in terms of \emph{case durations}, and (3) the log level, in 
terms of overall \emph{process workload}.

\vspace{-1em}
 \begin{table}
	\setlength{\tabcolsep}{3.5pt}
	\begin{tabular}{l rrrrrr }
		\toprule
		Attribute &  Original & $\epsilon=2.0$ & $\epsilon=1.5$  & $\epsilon=1.0$  & $\epsilon=0.5$  & $\epsilon=0.1$  \\
		\midrule
		Infection Suspected (fraction) & 0.81  & 0.75 & 0.69  & 0.67  &0.58   & 0.51\\
		Avg. Case Duration (days) & 28.47  & 36.93  &  7.95 & 37.77 & 37.16 &  34.2 \\
		Median Case Duration  (days) &  5.34 & 11.23 & 0.12  & 11.92  & 10.95  &  9.57 \\
		\bottomrule 
	\end{tabular}
\vspace{.7em}
	\caption{Sensitivity of attribute values to parameter $\epsilon$} 
	\label{tab:sensitivity_attribute}
\vspace{-3em}
\end{table}

\smallskip
\noindent \emph{Data attribute values:} At the event level, we compare the 
value distribution of data attributes in anonymised logs to the original 
distribution.
The Sepsis log primarily has attributes with boolean values. The quality of 
their value distributions is straightforward to quantify, i.e., by comparing 
the fraction of true values in an anonymised log $L'$ to the fraction in $L$. 
To illustrate the impact of the differential privacy parameter $\epsilon$ on 
attribute value quality, we assess the value distribution for the boolean 
attribute 
\emph{InfectionSuspected}.
As depicted in \autoref{tab:sensitivity_attribute}, the truth value of this 
attribute is true 
for 81\% of the cases in the original log. 

The anonymised distribution is 
reasonably preserved for 
the highest $\epsilon$ value, i.e., the least strict privacy guarantee. There, 
the distribution has 75\% true values. However, the accuracy of the 
distribution drops for stronger privacy guarantees, reaching almost full 
randomness for $\epsilon = 0.1$. This illustrates that the quality of attribute 
values 
can be preserved for certain privacy levels, but that it may be impacted for 
stricter settings. Note that, given that these results are obtained by 
anonymising individual values, the reduced quality for stronger privacy 
guarantees is inherently tied to the notion of differential privacy and is, 
therefore, independent of the specifics of the PRIPEL framework.

\medskip
\noindent \textit{Case duration.} 
Next, we investigate the accuracy of the case durations in the anonymised logs. 
Unlike the previously discussed quality of individual event attributes, the 
quality of case durations is influenced by all three steps of the framework. 
Therefore, when interpreting the results depicted in 
\autoref{tab:sensitivity_attribute}, it is important to consider that 
the maximal length of a trace is bound to 30 events in anonymised logs (due to 
the selection of parameter $n$), whereas the original log contains traces with 
up to 370 events. 
However, we can still observe longer case durations in the anonymised logs due 
to the added noise.  Additionally, in all scenarios, the average case duration 
is far higher than the median case duration. This indicates that the log 
contains several outliers in terms of longer case durations. All anonymised logs
reveal this insight.  We conclude that \emph{PRIPEL} preserves 
insights on the trace level, such as the duration of cases.

\begin{figure}[t]
	\vspace{-1em}
	\centering
	\includegraphics[width=.95\columnwidth]{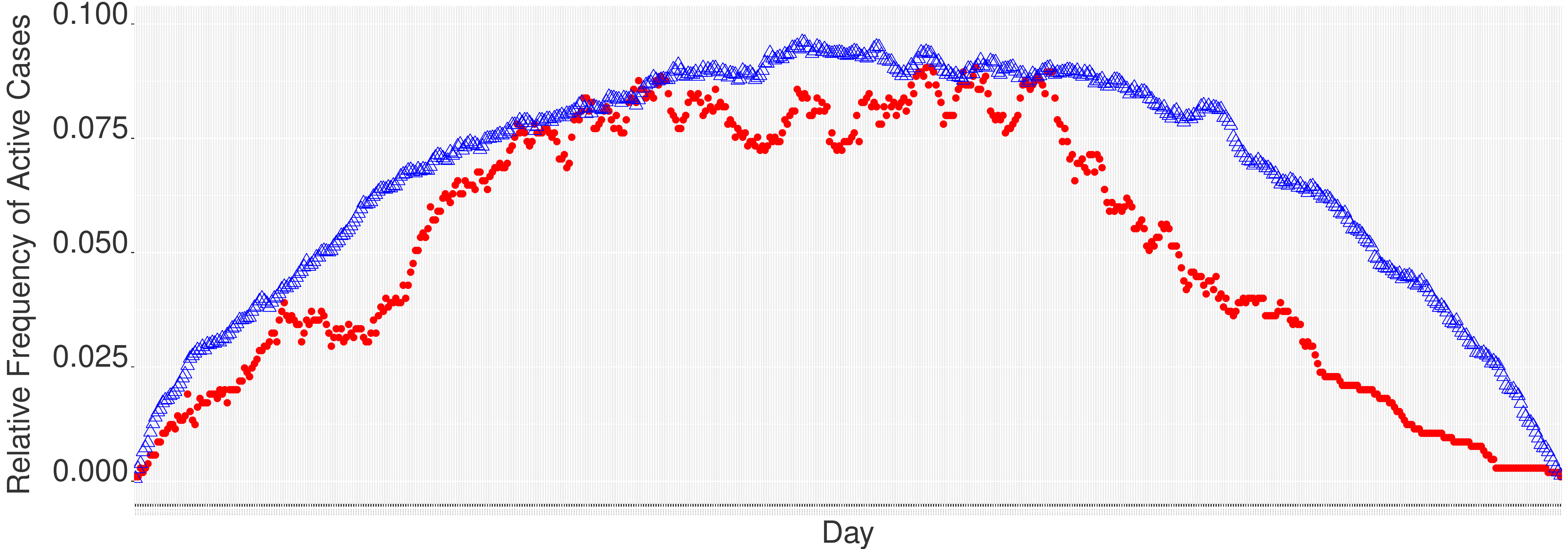}
	\caption{Active cases over time in original log (red) vs. anonymised 
	log (blue)}
	\label{fig:active_cases}
	\vspace{-1.7em}
\end{figure}

\medskip
\noindent \textit{Process workload.} Finally, at the log level, we consider the 
total workload of a process in terms of the number of cases that are active at 
any particular time. Given that anonymised event logs can have a considerably 
higher number of traces than the original log, we consider the progress of the 
relative number of active cases over time, as visualized in 
\autoref{fig:active_cases}. 
The red dots denote the original event log, while blue triangles represent the anonymised event log with $\epsilon=1.0$.

The figure clearly shows
that the general trend over time is sustained. However, 
the anonymised log shows a consistently higher workload than the original 
log. Furthermore, the variance over time is less extreme for the anonymised 
log. This shows that the necessary noise insertion smooths out some of the 
variability. Nevertheless, the results illustrate PRIPEL's ability to preserve 
utility for such a log-level process analysis.

\section{Discussion}
\label{sec:discussion}
With \emph{PRIPEL}, we introduced a framework that enables publishing of event 
logs that retain contextual information while guaranteeing differential 
privacy. As such, the anonymised event log can be used for rich process mining 
techniques that incorporate a fine-granular separation of classes of cases, 
without violating recent privacy regulations, such as the GDPR or CCPA.

While our general framework is generally applicable, the specific 
instantiations introduced earlier impose two assumptions on the event logs 
taken as input.

First, the employed notion for differential privacy assumes that any individual, such as a patient, is only 
represented in one case. 
To be able to guarantee differential privacy in contexts where this assumption may not hold, one can ensure that a single case exists per individual during the log extraction step, e.g., by limiting the 
selection of cases for which traces are derived or by 
constraining the time interval considered in the analysis. 
Alternatively, if the maximum number of cases per individual is known, the 
degree of noise introduced in the first step of the framework can be adjusted 
accordingly, by selecting the parameter $\epsilon$.
Finally, one may incorporate strategies that explicitly aim at adjusting 
differential privacy to handle multiple occurrences of individuals, such 
as~\cite{kartal2019differential}.

Second, we assume that all attributes can be anonymised independently. Hence, 
the usefulness of anonymised values or the degree of privacy may be reduced for 
strongly correlated attributes.
For instance, the independent anonymisation of the \emph{height} and \emph{age} of a child may result in improbable combinations.

Also, an attribute may represent a 
measurement that appears repeatedly in the trace, e.g., capturing the trend of 
a person's weight.
Since the measurements are inter-related, the values to be anonymised are not 
independent, so that the parallel composition of differential privacy is not 
applicable. 
In that case, one can employ notions of differential privacy as developed for 
streaming settings~\cite{dwork2010differential}.

Aside from these assumptions, we also acknowledge certain limitations related 
to our instantiation of the framework's steps. For instance, the approach chosen to determine the sensitivity of numerical attributes 
and timestamps is prone to outliers. Therefore, it might be necessary to reduce the number of outliers in an event log during pre-processing, in order to 
maintain the utility 
of the anonymised log. Yet, such limitations are inherent to any data anonymisation approach, since it has been shown that anonymisation reduces the utility of data~\cite{brickell2008cost}.
Another limitation relates to the applied trace variant query. For this query mechanism, the size of the 
anonymised log can differ drastically from the original log. This may diminish 
the utility of the log for certain analysis tasks, such as the identification 
of performance bottlenecks.

Finally, we highlight that the PRIPEL framework, and the notion of differential 
privacy in general, is particularly suited for analysis techniques that aim to 
aggregate or generalize over the traces in an (anonymised) event log. This 
means that the resulting event logs are suitable for, e.g., process discovery 
(e.g., by a directly-follows relation over all traces), 
log-level conformance checking (e.g., by a frequency 
distribution of deviations observed in all traces), 
process enhancement (e.g., by aggregate performance 
measures for activities), and predictive monitoring (e.g., by models that 
generalize the correlations observed between trace features and outcomes). 
However, the insertion of noise can lead to the inclusion of process behaviour 
that never occurred in the original log, which may lead to incorrect results 
when performing trace-level analysis, such as the establishment of alignments 
for a single case. If it is important to avoid such false positives, other 
anonymisation approaches, such as PRETSA~\cite{icpm/Fahrenkrog-Petersen19}, may 
be more suitable.

\vspace{-0.25em}
\section{Related Work}
\label{sec:relatedwork}

Privacy in process 
mining recently received a lot of 
attention~\cite{caise/Fahrenkrog-Petersen19,PikaWBHAR19}. The problem was 
raised 
in~\cite{MannhardtPO18}, noticing that most individuals might agree with 
the usage of their data for process improvement. However, the 
analysis of personal data for such a goal represents so-called 
secondary use, which is in violation of regulations such as the GDPR and CCPA.
Furthermore, in~\cite{saskia2019}, it was shown that 
even projections of event logs can lead to serious re-identification risks.

Several approaches have been proposed to address these privacy issues.
In~\cite{icpm/Fahrenkrog-Petersen19}, we proposed an algorithm to sanitize event logs for 
process discovery, which ensures $k$-anonymity and 
$t$-closeness.
Alternative 
approaches~\cite{burattin2015toward,rafiei2018ensuring} use cryptography to 
hide 
the confidential data in event logs. Other work focused on ensuring privacy 
for specific process mining tasks, by directly adapting analysis techniques. 
For instance, in~\cite{rafiei2019mining} a technique to ensure confidentiality in role mining 
was proposed, while~\cite{mannhardt2019privacy} introduced privacy-preserving 
queries to retrieve a directly-follows graph and the trace variants of a log. The work in~\cite{tillem2016privacy} 
uses encryption to calculate the output of the alpha miner in a 
privacy-preserving manner.
Other work considers process mining performed by multiple parties on an 
inter-organizational business process. 
In~\cite{liu2016towards}, an approach to generate a combined process model for 
such a business process was proposed. Similarly, \cite{corr/abs-1912-01855} 
introduces an approach based on secure multi-party computation to answer 
queries relating the business process, such as the directly-follows query.

\vspace{-0.25em}
\section{Conclusion}
\label{sec:conclusion}

In this paper, we introduced {PRIPEL}, a framework to publish anonymised 
event logs that incorporates contextual information while guaranteeing differential privacy. 
In particular, PRIPEL ensures differential privacy on the basis of individual cases, rather than on an entire event log. We achieved this by exploiting the maxim of parallel composition.
By applying a prototypical implementation on a real-world event log, we illustrate that the utility of anonymised event logs is preserved for various types of analysis involving contextual information.

By incorporating contextual information, for the first time, {PRIPEL} offers 
the use of rich process mining techniques in a privacy-preserving manner. In 
particular, anonymised event logs are now suitable for analysis techniques that incorporate 
a fine-granular separation of cases based on contextual information. In 
future work, we intend to further explore the impact that strongly correlated attributes have on the provided privacy guarantees. 
In addition, we aim to incorporate the handling of ongoing cases in the {PRIPEL} framework.

\medskip
\noindent
\textbf{Acknowledgements}
This work was partly supported by the Alexander von Humboldt Foundation. 

\bibliographystyle{splncs04}

\begin{thebibliography}{10}
\providecommand{\url}[1]{\texttt{#1}}
\providecommand{\urlprefix}{URL }
\providecommand{\doi}[1]{https://doi.org/#1}

\bibitem{augusto2018automated}
Augusto, A., Conforti, R., Dumas, M., La~Rosa, M., Maggi, F.M., Marrella, A.,
  Mecella, M., Soo, A.: Automated discovery of process models from event logs:
  Review and benchmark. IEEE Transactions on Knowledge and Data Engineering
  \textbf{31}(4),  686--705 (2018)

\bibitem{berti2019process}
Berti, A., van Zelst, S.J., van~der Aalst, W.: Process mining for python
  (pm4py): bridging the gap between process-and data science. arXiv preprint
  arXiv:1905.06169  (2019)

\bibitem{brickell2008cost}
Brickell, J., Shmatikov, V.: The cost of privacy: destruction of data-mining
  utility in anonymized data publishing. In: Proceedings of the 14th ACM SIGKDD
  international conference on Knowledge discovery and data mining. pp. 70--78
  (2008)

\bibitem{burattin2015toward}
Burattin, A., Conti, M., Turato, D.: Toward an anonymous process mining. In:
  2015 3rd International Conference on Future Internet of Things and Cloud. pp.
  58--63. IEEE (2015)

\bibitem{dwork2006calibrating}
Dwork, C., McSherry, F., Nissim, K., Smith, A.: Calibrating noise to
  sensitivity in private data analysis. In: Theory of cryptography conference.
  pp. 265--284. Springer (2006)

\bibitem{dwork2010differential}
Dwork, C., Naor, M., Pitassi, T., Rothblum, G.N.: Differential privacy under
  continual observation. In: Proceedings of the forty-second ACM symposium on
  Theory of computing. pp. 715--724 (2010)

\bibitem{dwork2014algorithmic}
Dwork, C., Roth, A., et~al.: The algorithmic foundations of differential
  privacy. Foundations and Trends{\textregistered} in Theoretical Computer
  Science  \textbf{9}(3--4),  211--407 (2014)

\bibitem{corr/abs-1912-01855}
Elkoumy, G., Fahrenkrog{-}Petersen, S.A., Dumas, M., Laud, P., Pankova, A.,
  Weidlich, M.: Secure multi-party computation for inter-organizational process
  mining. CoRR  \textbf{abs/1912.01855} (2019),
  \url{http://arxiv.org/abs/1912.01855}

\bibitem{erlingsson2014rappor}
Erlingsson, {\'U}., Pihur, V., Korolova, A.: Rappor: Randomized aggregatable
  privacy-preserving ordinal response. In: Proceedings of the 2014 ACM SIGSAC
  conference on computer and communications security. pp. 1054--1067. ACM
  (2014)

\bibitem{wp216}
of~the EU~Commission, A..D.P.W.P.: Opinion 05/2014 on anonymisation techniques
  (2014)

\bibitem{caise/Fahrenkrog-Petersen19}
Fahrenkrog{-}Petersen, S.A.: Providing privacy guarantees in process mining.
  In: Proceedings of the Doctoral Consortium Papers Presented at the 31st
  International Conference on Advanced Information Systems Engineering (CAiSE
  2019), Rome, Italy, June 3-7, 2019. pp. 23--30 (2019),
  \url{http://ceur-ws.org/Vol-2370/paper-03.pdf}

\bibitem{icpm/Fahrenkrog-Petersen19}
Fahrenkrog{-}Petersen, S.A., van~der Aa, H., Weidlich, M.: {PRETSA:} event log
  sanitization for privacy-aware process discovery. In: International
  Conference on Process Mining, {ICPM} 2019, Aachen, Germany, June 24-26, 2019.
  pp.~1--8 (2019). \doi{10.1109/ICPM.2019.00012},
  \url{https://doi.org/10.1109/ICPM.2019.00012}

\bibitem{garfinkel2015identification}
Garfinkel, S.L.: De-identification of personal information. National institute
  of standards and technology  (2015)

\bibitem{hintze2018viewing}
Hintze, M.: Viewing the gdpr through a de-identification lens: a tool for
  compliance, clarification, and consistency. International Data Privacy Law
  \textbf{8}(1),  86--101 (2018)

\bibitem{holohan2019diffprivlib}
Holohan, N., Braghin, S., Mac~Aonghusa, P., Levacher, K.: Diffprivlib: The ibm
  differential privacy library. arXiv preprint arXiv:1907.02444  (2019)

\bibitem{holohan2017optimal}
Holohan, N., Leith, D.J., Mason, O.: Optimal differentially private mechanisms
  for randomised response. IEEE Transactions on Information Forensics and
  Security  \textbf{12}(11),  2726--2735 (2017)

\bibitem{kartal2019differential}
Kartal, H.B., Liu, X., Li, X.B.: Differential privacy for the vast majority.
  ACM Transactions on Management Information Systems (TMIS)  \textbf{10}(2),
  1--15 (2019)

\bibitem{kasiviswanathan2011can}
Kasiviswanathan, S.P., Lee, H.K., Nissim, K., Raskhodnikova, S., Smith, A.:
  What can we learn privately? SIAM Journal on Computing  \textbf{40}(3),
  793--826 (2011)

\bibitem{kessler2019sap}
Kessler, S., Hoff, J., Freytag, J.C.: Sap hana goes private: from privacy
  research to privacy aware enterprise analytics. Proceedings of the VLDB
  Endowment  \textbf{12}(12),  1998--2009 (2019)

\bibitem{lefevre2005incognito}
LeFevre, K., DeWitt, D.J., Ramakrishnan, R.: Incognito: Efficient full-domain
  k-anonymity. In: Proceedings of the 2005 ACM SIGMOD international conference
  on Management of data. pp. 49--60 (2005)

\bibitem{levenshtein1966binary}
Levenshtein, V.I.: Binary codes capable of correcting deletions, insertions,
  and reversals. In: Soviet physics doklady. vol.~10, pp. 707--710 (1966)

\bibitem{liu2016towards}
Liu, C., Duan, H., Qingtian, Z., Zhou, M., Lu, F., Cheng, J.: Towards
  comprehensive support for privacy preservation cross-organization business
  process mining. IEEE Transactions on Services Computing  (2016)

\bibitem{maggi2014predictive}
Maggi, F.M., Di~Francescomarino, C., Dumas, M., Ghidini, C.: Predictive
  monitoring of business processes. In: International conference on advanced
  information systems engineering. pp. 457--472. Springer (2014)

\bibitem{mannhardt2016sepsis}
Mannhardt, F.: Sepsis cases-event log. Eindhoven University of Technology.
  Dataset pp. 227--228 (2016)

\bibitem{mannhardt2019privacy}
Mannhardt, F., Koschmider, A., Baracaldo, N., Weidlich, M., Michael, J.:
  Privacy-preserving process mining. Business \& Information Systems
  Engineering  \textbf{61}(5),  595--614 (2019)

\bibitem{MannhardtPO18}
Mannhardt, F., Petersen, S.A., Oliveira, M.F.: Privacy challenges for process
  mining in human-centered industrial environments. In: 14th International
  Conference on Intelligent Environments, {IE} 2018, Roma, Italy, June 25-28,
  2018. pp. 64--71 (2018). \doi{10.1109/IE.2018.00017},
  \url{https://doi.org/10.1109/IE.2018.00017}

\bibitem{mcsherry2007mechanism}
McSherry, F., Talwar, K.: Mechanism design via differential privacy. In: 48th
  Annual IEEE Symposium on Foundations of Computer Science (FOCS'07). pp.
  94--103. IEEE (2007)

\bibitem{mcsherry2009privacy}
McSherry, F.D.: Privacy integrated queries: an extensible platform for
  privacy-preserving data analysis. In: Proceedings of the 2009 ACM SIGMOD
  International Conference on Management of data. pp. 19--30. ACM (2009)

\bibitem{PikaWBHAR19}
Pika, A., Wynn, M.T., Budiono, S., ter Hofstede, A.H., van~der Aalst, W.M.,
  Reijers, H.A.: Privacy-preserving process mining in healthcare. vol.~17,
  p.~1612. Multidisciplinary Digital Publishing Institute (2020)

\bibitem{rafiei2019mining}
Rafiei, M., van~der Aalst, W.M.: Mining roles from event logs while preserving
  privacy. In: International Conference on Business Process Management
  Workshops. pp. 676--689. Springer (2019)

\bibitem{rafiei2018ensuring}
Rafiei, M., von Waldthausen, L., van~der Aalst, W.M.: Ensuring confidentiality
  in process mining. In: SIMPDA. pp. 3--17 (2018)

\bibitem{dp2017learning}
Team, D., et~al.: Learning with privacy at scale. Online at:
  https://machinelearning. apple.
  com/2017/12/06/learning-with-privacy-at-scale. html  (2017)

\bibitem{tillem2016privacy}
Tillem, G., Erkin, Z., Lagendijk, R.L.: Privacy-preserving alpha algorithm for
  software analysis. In: 37th WIC Symposium on Information Theory in the
  Benelux/6th WIC/IEEE SP Symposium on Information Theory and Signal Processing
  in the Benelux (2016)

\bibitem{van2011process}
Van Der~Aalst, W., Adriansyah, A., De~Medeiros, A.K.A., Arcieri, F., Baier, T.,
  Blickle, T., Bose, J.C., Van Den~Brand, P., Brandtjen, R., Buijs, J., et~al.:
  Process mining manifesto. In: International Conference on Business Process
  Management. pp. 169--194. Springer (2011)

\bibitem{voigt2017eu}
Voigt, P., Von~dem Bussche, A.: The eu general data protection regulation
  (gdpr). A Practical Guide, 1st Ed., Cham: Springer International Publishing
  (2017)

\bibitem{saskia2019}
von Voigt, S.N., Fahrenkrog-Petersen, S.A., Janssen, D., Koschmider, A.,
  Tschorsch, F., Mannhardt, F., Landsiedel, O., Weidlich, M.: Quantifying the
  re-identification risk of event logs for process mining. In: Accepted for
  publication in CAiSE 2020 (2020)

\bibitem{warner1965randomized}
Warner, S.L.: Randomized response: A survey technique for eliminating evasive
  answer bias. Journal of the American Statistical Association
  \textbf{60}(309),  63--69 (1965)

\bibitem{zhang2006outsourcing}
Zhang, J., Borisov, N., Yurcik, W.: Outsourcing security analysis with
  anonymized logs. In: 2006 Securecomm and Workshops. pp.~1--9. IEEE (2006)

\end{thebibliography}

\end{document}